\documentclass{PoS}
\usepackage{graphicx}
\usepackage{epsfig}
\usepackage{subfigure}
\usepackage{amsmath}
\title{Search for exotic charmonium states}

\ShortTitle{Exotic states}

\author{\speaker{Elisabetta Prencipe}\thanks{(previously addressed at JGU University of Mainz, Germany)}\\
        representing the $BABAR$ and the Belle Collaborations.\\
        Forschungszentrum J\"ulich, 52428 J\"ulich, Germany\\
        E-mail: \email{e.prencipe@fz-juelich.de}}

\abstract{ 
One of the most intriguing puzzles in hadron spectroscopy are the numerous charmonium-like states observed in the last decade, including charged states that are manifestly exotic. The $BABAR$  and Belle experiments have extensively studied those in B meson decays, initial state radiation processes and two photon reactions.  We can question what we have understood after 11 year search in this field,  and try to combine results to conclude on what these new unpredicted resonant states are, and how they can be accommodated in the theory. Big effort has been made from theoretical and experimental point of view, as the potential models unlikely explain the presence of so many enhancements, for mass values above the $D \bar D$ threshold. In this report the $BABAR$ and Belle results of the two invariant mass systems  of $J/\psi \phi$ and $J/\psi \omega$  are put in comparison in a search for non-conventional charmonium states. This involves the study of the systems of $J/\psi K^+ K^-$ and $J/\psi \pi^+ \pi^- \pi^0$, respectively. There are strong theoretical arguments in favor of the presence of hybrids or exotic states, in those invariant mass distributions. 

Remarks on these data analyses are given, based on the $BABAR$  and Belle experimental results.}

\FullConference{XIIth International Conference on Heavy Quarks $\&$ Leptons 2014,\\
		25-29 August 2014\\
		Schloss Waldthausen, Mainz, Germany}

\begin{document}

\section{Introduction}
Since the discovery of the $X(3872)$\cite{ref1} many new resonant states have been observed, revealing a spectrum too rich to be uniquely described  by potential models\cite{ref2}. Some of these states  were predicted and recently observed; some others were  predicted but have never been found; some others were seen, but they were never predicted by any theoretical model. In some cases these enhancements have been controversial, as several of those have not been confirmed by all experiments. We use to name as X, Y, Z the unpredicted states whose nature is still unclear, due mostly to the inability to assign quantum numbers.  The logic question might be what we have understood after 11 years of these unexpected findings. In trying to answer to this question, in this report the experimental results related to the study of two invariant mass distributions are compared, using the $BABAR$ and the Belle data sets.

The QCD spectrum is much richer than that of the naive quark model, as the gluons, which mediate the strong force between quarks, can also act as principal components of entirely new types of hadrons: glueballs, hybrids, etc. In searching for these new forms of aggregation of matter, we analyze systems where the production of new resonant states can be enhanced: B color-suppressed decays, $\gamma \gamma$ interaction processes, ISR processes, and double-charmonium production.   

The invariant mass systems of $J/\psi \omega$ and $J/\psi \phi$ have been analyzed in B decays and two-photon production processes. These two invariant mass distributions show similarities: both constituent particles are vector states, the invariant mass range is larger than the $D \bar D$ threshold, and the production of exotics is potentially allowed  in both cases. The invariant mass distribution of $J/\psi \omega$ is known for the search of the $X(3872)$ and the $X(3915)$; the invariant mass distribution of $J/\psi \phi$ has recently gained attention because of the controversial claims of the $X(4140)$ and the $X(4270)$.

\section{Remarks on the analysis of the invariant mass $J/\psi \omega$}

The invariant mass system of $J/\psi \omega$ was analyzed in B decays (in the process $B \rightarrow J/\psi \omega K$) and via $\gamma \gamma$ interactions. The invariant mass distribution $J/\psi \omega$ is definitively not PHSP distributed. Several resonant states have been observed in the mass range available for the system of  $J/\psi \omega$. We know them as the $X(3940)$, the $X(3915)$, the $Y(3940)$ and the $Z(3930)$: they have been observed in different decay modes, and they have been measured with similar mass values. The logic question is if they can be identified as the same resonance, how they can be accommodated in the theory and if there are predictions in the potential models for these states. The answer to these questions comes from experimental observations and from the results of the angular distribution analyses.
 
The $X(3915)$ resonance, decaying to the $J/\psi \omega$ final state, was first observed by the Belle Collaboration in two-photon collisions\cite{belle1}. Another resonance, dubbed $Y(3940)$, has been observed in the $B \rightarrow J/\psi \omega K$ process by Belle\cite{belle2, belle5}, then confirmed by $BABAR$\cite {belle3, belle4}. The mass measurement for the $Y(3940)$ observed by Belle, is consistent with that of the $X(3915)$ observed by $BABAR$. Thus, the same particle, with a mass of about 3915 MeV/c$^2$, may have been observed in two distinct production processes. The $Z(3930)$ resonance has been discovered in the $\gamma \gamma \rightarrow D \bar D$ process\cite{belle6, belle7}. Its interpretation as the $\chi_{c2}(2P)$, the first radial excitation of the $^3P_2$ charmonium ground state, is commonly accepted\cite{belle8}. Interpretation of the $X(3915)$ as the
$\chi_{c0}(2P)$\cite{belle9} or $\chi_{c2}(2P)$ state\cite{belle10} has been suggested. The latter implies that the $X(3915)$ and $Z(3930)$ are the same
particle, observed in different decay modes. However, the product of the two-photon width times the decay branching fraction (BF) for the $X(3915)$ reported by Belle\cite{belle5} is unexpectedly large compared to other excited $c \bar c$ states\cite{belle8}. Interpretation of the $X(3915)$ in the framework of molecular models has also been proposed\cite{belle11}.
\begin{figure}[htb]
\centering
\mbox{ 
{\scalebox{0.42}{\includegraphics{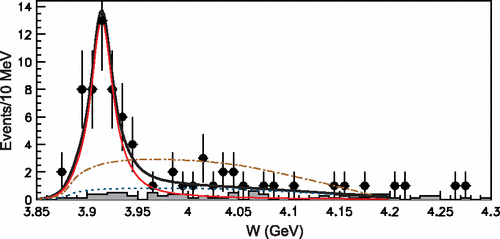}}} \quad 
{\scalebox{0.30}{\includegraphics{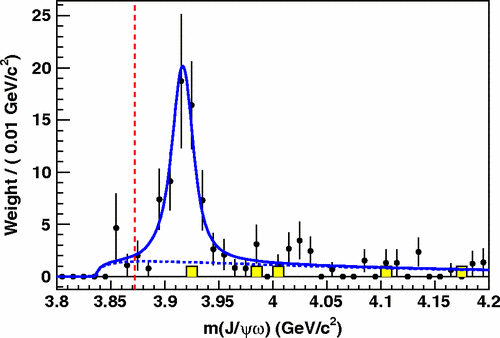}}}
}
\caption{\label{Fig1-prencipe} Invariant mass distributions of $J/\psi \omega$ produced in the process $e^+ e^- \rightarrow J/\psi \omega$ via $\gamma \gamma$ fusion at Belle (left) and BaBar (right).}
\end{figure}

$BABAR$ and Belle have searched for the $X(3915)$, already observed from both in B decays,  in the process $e^+e^- \rightarrow J/\psi \omega$ via $\gamma \gamma$ fusion\cite{ref3, ref4}, and they have found excellent agreement in the mass measurement of this resonance. In Fig.~\ref{Fig1-prencipe} the plots published from the 2 experiments for the $X(3915)$ produced in two-photon interactions are shown. Results are reported in Table~\ref{table1-prencipe}.

The $X(3872)$ and the $X(3915)$ were found in their decay to $J/\psi \omega$ in $B$ decays; however, no evidence for the signal of the $X(3872)$ was found in $J/\psi \omega$ via two-photon interactions.

The angular analysis performed by $BABAR$\cite{babarX} to understand the nature of the X(3915) allowed to fix its quantum numbers: it is very likely a state with $J^P$ = $0^+$. Logically we can think that the $X(3915)$ is the missing state $\chi_{c0} (2P)$ of the charmonium spectrum. However, the mass difference between the two states, $X(3915)$ and $Z(3930)$, identified as the $\chi_{c0} (2P)$ and $\chi_{c1} (2P)$ respectively, is found to be $\sim$10 MeV, for an excitation from $J=0$ to $J=2$. Just for comparison, the equivalent difference between the states $\chi_{cJ}(1P)$, with $J =0,2$, is roughly 142 MeV. This poses into question the interpretation of the $X(3915)$ as $\chi_{c0} (2P)$, or it brings us to question the model. The mass and width of the $X(3915)$, found in $\gamma \gamma$ fusion, are consistent with the parameters published by $BABAR$ for the state decaying to $J/\psi \omega$ in B decays; therefore we can conclude that it is the same particle. These values are also slightly consistent with the parameters of the state called $Y(3940)$ that Belle observed in 2005, in the decay $B \rightarrow J/\psi \omega K$, decaying to $J/\psi \omega$.  A summary of these conclusions can be find in Table~\ref{table1-prencipe}.
\begin{table}
\centering
\caption{Summary of combined results for X(3940), X(3915), Y(3940) and Z(3930)}
\label{table1-prencipe}
\begin{tabular}{lllll}\hline
Resonance &Mass (MeV/c$^2$)& Width (MeV)&J$^{\rm PC}$&Interpretation\\ \hline
X(3940)  & 3942 $\pm$ 9 &  37$\rm ^{+27}_{-17}$& &\\  
Y(3940)/X(3915) & 3918.4 $\pm$ 1.9& 20 $\pm$  5& 0$^+$ & $\chi_{c0}(2P)$?\\ 
Z(3930)  & 3927.2 $\pm$ 2.6& 24 $\pm$ 6& 2$^{++}$& $\chi_{c2}(2P)$\\ \hline
\end{tabular} 
\end{table}

\section{Remarks on the analysis of the invariant mass $J/\psi \phi$}
$Strangeness$ in charmonium seems a sector still to be exploited.
While resonant structures like  the X(3872) have been seen in $B \rightarrow X K, X \rightarrow J/\psi~ \pi^+ \pi^-$, or like Y(4260) by investigating the process $e^+e^- \rightarrow \gamma_{ISR}X$, $X \rightarrow J/\psi \pi^+ \pi^-$\cite{X3872, Y4260babar, ref1}, no indication of new states has been observed in the  $J/\psi~ K^+ K^-$ invariant mass system, until the paper quoted in Ref.~\cite{kai} highlighted the possibility of a couple of resonant states, decaying to $J/\psi\phi$, with $\phi \rightarrow K^+ K^-$ and $J/\psi \rightarrow \mu^+ \mu^-$. These observations are nowaday controversial.

 The rare decay $B \rightarrow J/ \psi \phi K$ is interesting because it is a promising place to search for new resonances, as it proceeds, at quark level, via the weak transition $b \rightarrow c \bar c s$. It could be a quasi-two-body decay, $B \rightarrow X_g K$, with $X_g \rightarrow J/ \psi \phi$, where $X_g$ = $\lvert g c \bar c s \bar s>$, with gluonic contribution ($g$).

$BABAR$ has looked for possible resonant states decaying to two mesons: $J/ \psi$, and  another meson  with strange $s$-quark content. The present $BABAR$ analysis describes the case of $X_g \rightarrow J/\psi \phi$, $\phi \rightarrow K^{+}K^{-}$ in $B$ decays. $BABAR$ looked also for exotic charged states, such as $Z \rightarrow J/\psi K^+$. $J/\psi$ and $\phi$ are 2 vector states, so non-parametrizable polarization effects can be shown in the dynamics of their interaction. Thus, more complications can arise compared to the phase-space (PHSP) model. Exotic quantum number combinations are theoretically allowed in this case. Predictions for hybrids come mainly from calculations based on the bag model, flux tube model, constituent gluon model and recently, with increasing precision, from Lattice QCD. 

The analysis $B^+ \rightarrow J/\psi K^+K^- K^+$,  $B^0 \rightarrow J/\psi K^+K^- K^0_S$, $B^+ \rightarrow J/\psi \phi  K^+$ and $B^0 \rightarrow J/\psi \phi  K^0_S$ are performed by $BABAR$ using 469 million pairs. With $B^+$ we will imply in the text also the charged conjugate $B^-$. The $\phi(1020)$ signal region is selected  in the mass range [1.004; 1.034] GeV/c$^{\rm 2}$. $J/\psi$ is recontructed through its decays to $e^+e^-$ and $\mu^+ \mu^-$, then it is mass constrained.  
An unbinned  maximum likelihood fit is performed  to extract the yield and calculate the BFs. Detailed explanation of these calculations are reported in Ref.~\cite{elisabetta}, together with the relevant discussion for the non-resonant $K^+K^-$ contribution to the BF of $B \rightarrow J/\psi KKK$ and systematic uncertainty calculation. For the first time the non-resonant $K^+K^-$ contribution to the BF of $B \rightarrow J/\psi K K  K$ is observed. Here we report only the relevant information for the analysis of the three invariant mass distributions: $J/\psi \phi$, $J/\psi K$, $KKK$, for both charged and neutral B samples.

$BABAR$ has searched for the resonant states reported by the CDF Collaboration in the
$J/\psi \phi$ mass spectrum. The masses and the widths in the fit are fixed to values according to Ref.~\cite{kai}. 
\begin{figure}[htb]
\centering
\mbox{
{\scalebox{0.24}{\includegraphics{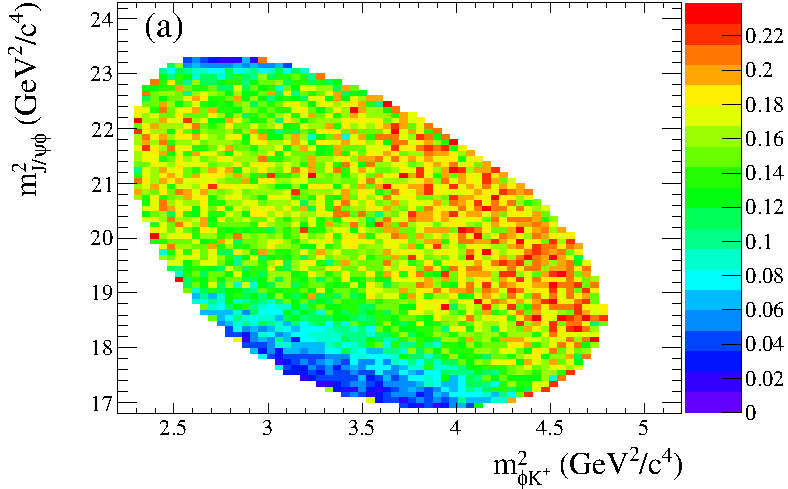}}} \quad 
{\scalebox{0.24}{\includegraphics{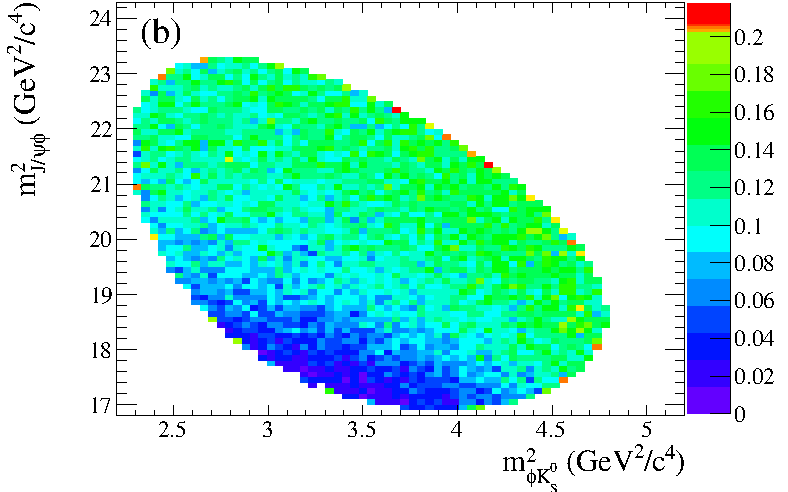}}} \quad 
{\scalebox{0.185}{\includegraphics{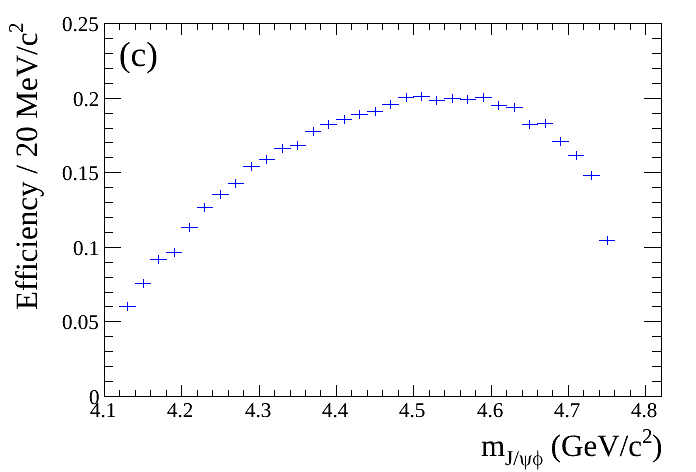}}}
}
\caption{\label{Fig2-prencipe} $BABAR$ simulations: efficiency distribution on the Dalitz plot for (a) $B^+ \rightarrow J/\psi \phi K^+$ and (b) $B^0 \rightarrow J/\psi \phi K^0_S$. (c) Average efficiency distribution as a function of the $J/\psi \phi$ mass for $B^+ \rightarrow J/\psi \phi K^+$.}
\end{figure}

\begin{figure}[htb]
\centering
\mbox{
{\scalebox{0.195}{\includegraphics{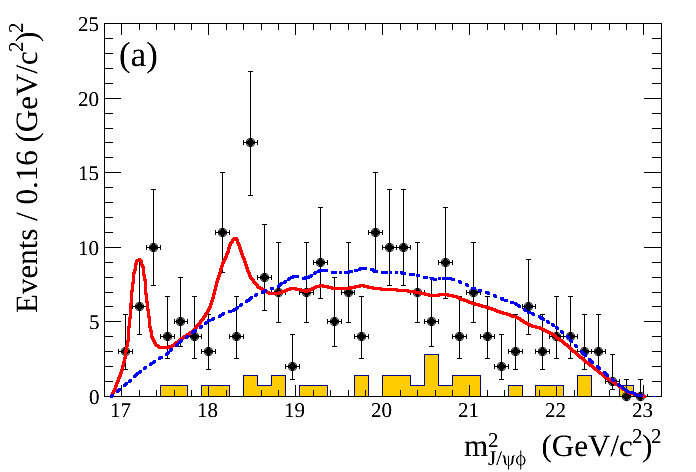}}} \quad
{\scalebox{0.195}{\includegraphics{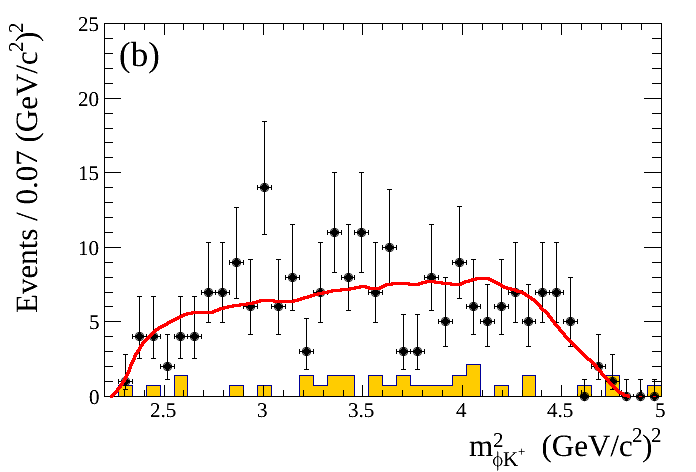}}} \quad
{\scalebox{0.195}{\includegraphics{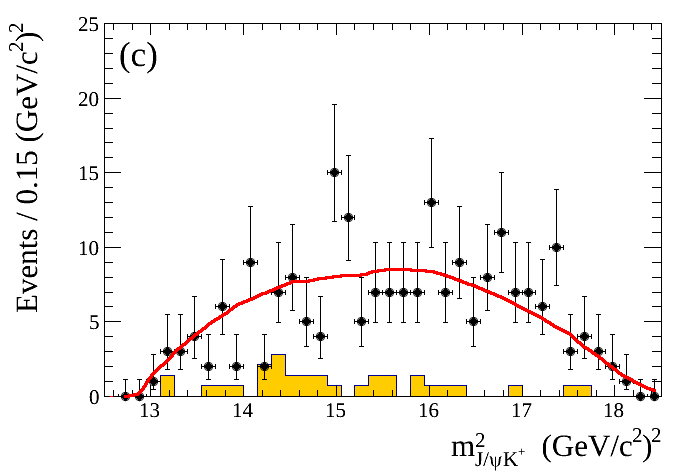}}}
}
\caption{\label{Fig3-prencipe} $BABAR$ data: Dalitz plot projections for $B^+ \rightarrow J/\psi \phi K^+$ on (a) $m^2_{J/\psi \phi}$, (b) $m^2_{\phi K^+}$, and (c) $m^2_{J/\psi K^+}$. The continuous (red) curves are the results from fit model performed including the $X(4140)$ and $X(4270)$ resonances. The dashed (blue) curve in (a)  indicates the projection for fit model  with no resonances included in the fit. The shaded (yellow) histograms indicate the evaluated background.}
\end{figure}
\begin{figure}[htb]
\centering
\mbox{
{\scalebox{0.24}{\includegraphics{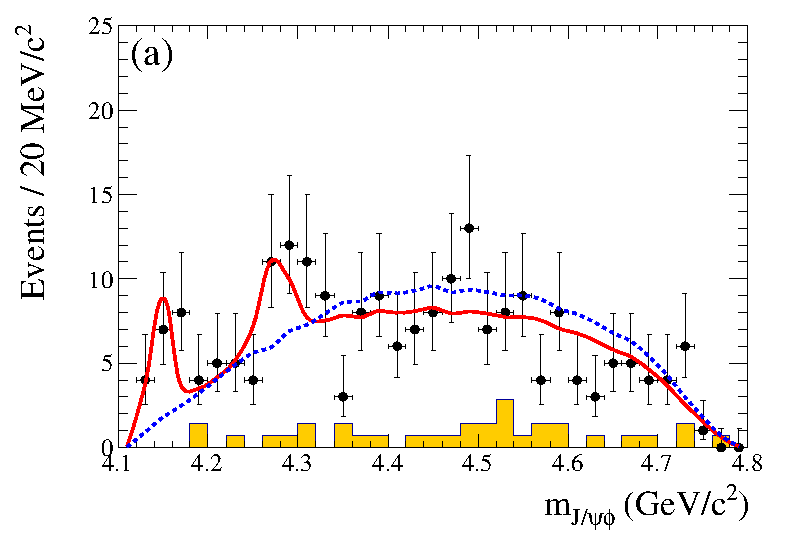}}} \quad
{\scalebox{0.24}{\includegraphics{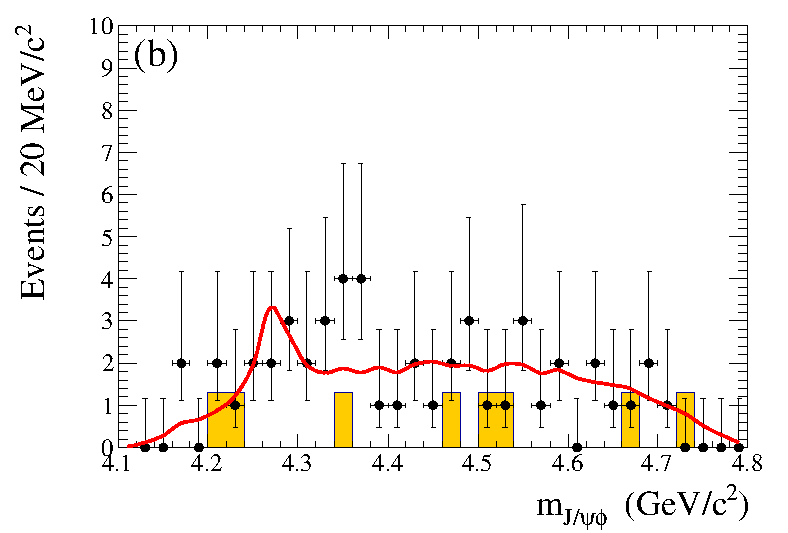}}} \quad
{\scalebox{0.24}{\includegraphics{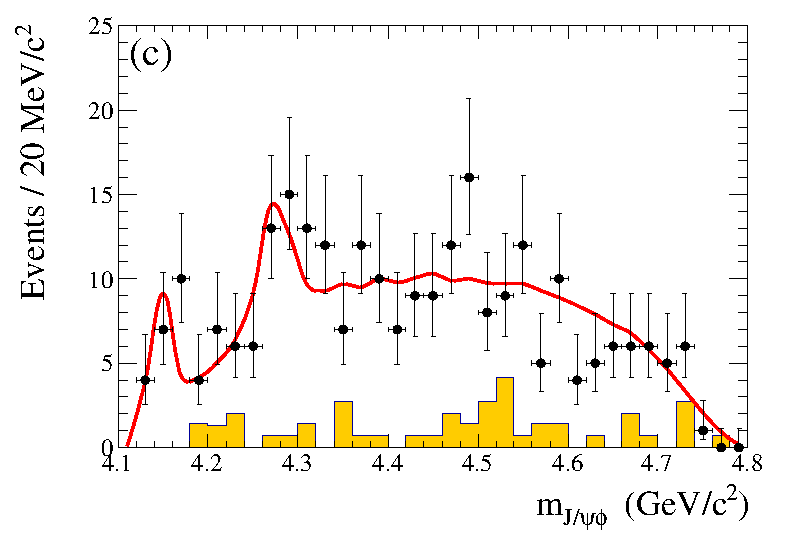}}}
}
\caption{\label{Fig3-babar} $BABAR$ data: projections on the $J/\psi \phi$ mass spectrum from the Dalitz plot fit with the $X(4140)$ and the $X(4270)$ resonances for the (a) $B^+$, (b) $B^0$, and (c) combined $B^+$ and  $B^0$ data samples. The continuous (red) curves result from the fit; the dashed (blue) curve in (a)  indicates the projection for fit model D, with no resonances. The shaded (yellow) histograms show the estimated background contributions.}
\end{figure}

A significant efficiency decrease at low $J/\psi \phi$ mass (see Fig.~\ref{Fig2-prencipe}) is observed by $BABAR$, due to the inability to reconstruct slow kaons in the laboratory frame, as a result of energy loss in the beampipe and SVT material. An unbinned maximum likelihood fit to the channel $B^+ \rightarrow J/\psi \phi K^+$ is performed, modeling the resonances with an inchoerent sum of two S-wave relativistic Breit-Wigner (BW) functions with parameters fixed to the CDF values~\cite{kai}. A non-resonant contribution is described according to PHSP.  The decay of a pseudoscalar meson to two vector states contains high spin contributions which could generate non-uniform angular distributions. However, due to the limited data sample (212 yield for $B^+$ and 50 for $B^0$, in the signal area, respectively) such angular terms have been not included, and the assumption is that the resonances decay isotropically. The fit function is  weighted by the inverse of the two-dimentional efficiency computed on the Dalitz plots (see the continuous red curve in Fig.~\ref{Fig3-prencipe} and Fig.~\ref{Fig3-babar}). $BABAR$ has perform the fits using models with two resonances, one resonance, and no resonances. All models provide a reasonably good description of the data, with $\chi^2$ probability larger than 5$\%$. The following corrected-estimates for the fractions for $B^+$ are then obtained, where the central values of mass and width of the two resonances are fixed to the values recently published by CDF\cite{kai} (Eq.~3.1) and CMS~\cite{cms} (Eq.~3.2), respectively:\\
\begin{eqnarray}
f_{X(4140)} = (9.2 \pm 3.3 \pm 4.7)\%,   f_{X(4270)} = (10.6 \pm 4.8 \pm 7.1) \%,  \end{eqnarray} 
\begin{eqnarray}
f_{X(4140)} = (13.2 \pm 3.8 \pm 6.8)\%,  f_{X(4270)} = (10.9 \pm 5.2 \pm 7.3) \%.
\end{eqnarray}

These values are consistent with each others within the uncertainties.
For comparison, CMS reported a fraction of $0.10 \pm 0.03$ for the X(4140), compatible with CDF, LHCb and our values within the uncertainties. CMS could not determine reliably the significance of the second structure X(4270) due to possible reflections of two-body decays. A significance smaller than 2$\sigma$ is found for the 2 peaks, within systematic uncertainties, by $BABAR$. Using the Feldman-Cousins method\cite{FC}, we obtain the ULs at 90\% CL: \\
$BF(B^+ \rightarrow X(4140)K^+)\times BF(X(4140) \rightarrow J/\psi \phi)/BF(B^+ \rightarrow J/\psi \phi K^+) < 0.135$, and \\
$BF(B^+ \rightarrow X(4270)K^+)\times BF(X(4270)\rightarrow J/\psi \phi)$/$BF$$(B^+ \rightarrow J/\psi \phi K^+) <$ $0.184$.

The $X(4140)$ limit may be compared with the CDF measurement of $0.149\pm 0.039\pm 0.024$~\cite{kai} and the LHCb limit of 0.07~\cite{LHCb}. The $X(4270)$ limit may be compared with the LHCb limit of 0.08. A detailed description of all BFs and ULs shortly introduced in this report is provided  in Ref.~\cite{elisabetta}: this work has been recently submitted to PRD.

\begin{figure}[htb]
\centering
\mbox{
{\scalebox{0.23}{\includegraphics{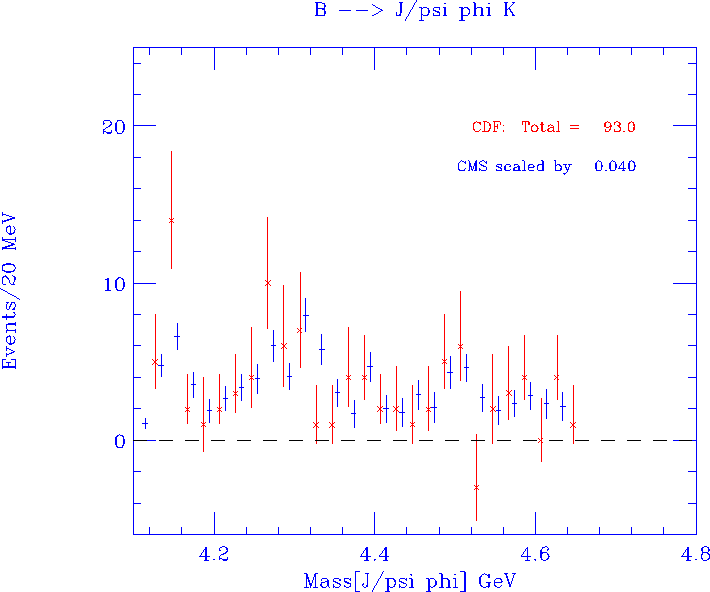}}} \quad 
{\scalebox{0.23}{\includegraphics{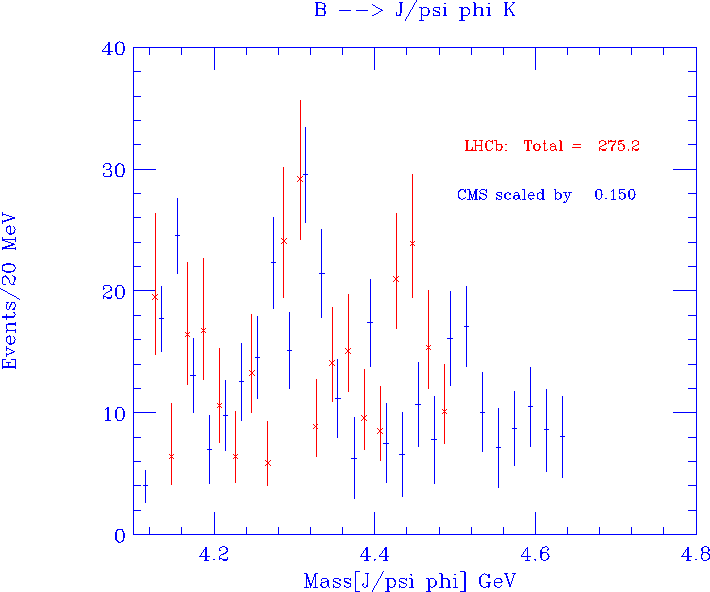}}}\quad  
{\scalebox{0.23}{\includegraphics{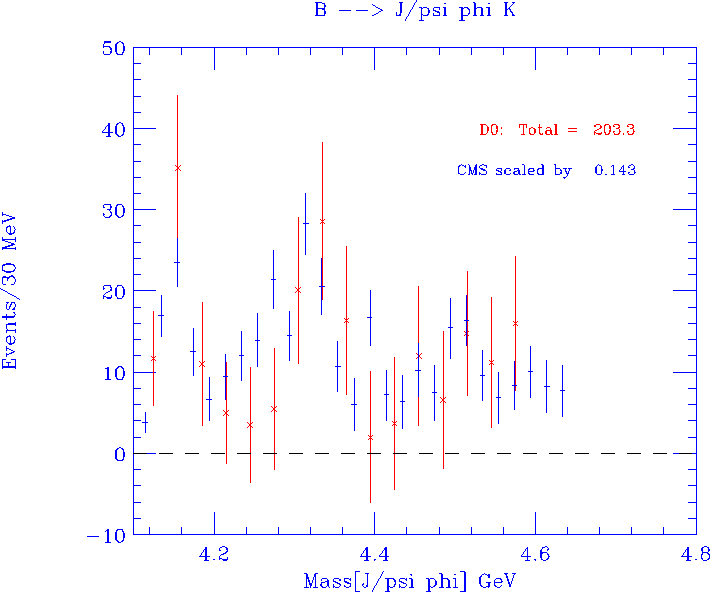}}}
}
\caption{\label{Fig4-prencipe} $J/\psi \phi$ invariant mass distributions for the comparison between the CDF and CMS data (left); LHCb and CMS data (center); D0 and CMS data (right). Informations are taken from the references ~\cite{kai, LHCb, cms, d0}. Data are rebinned, background-subtracted and properly scaled as indicated in the labels, for the comparison.}
\end{figure}

\begin{figure}[htb]
\centering
\mbox{ 
{\scalebox{0.22}{\includegraphics{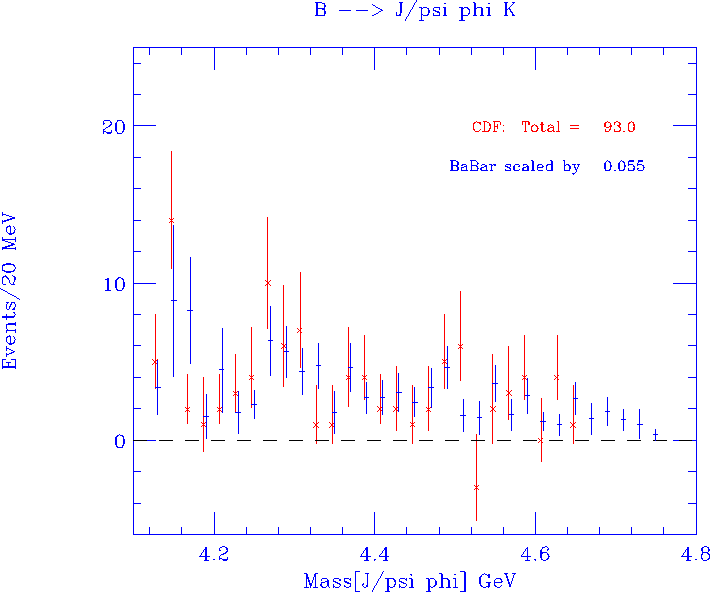}}} \quad
{\scalebox{0.22}{\includegraphics{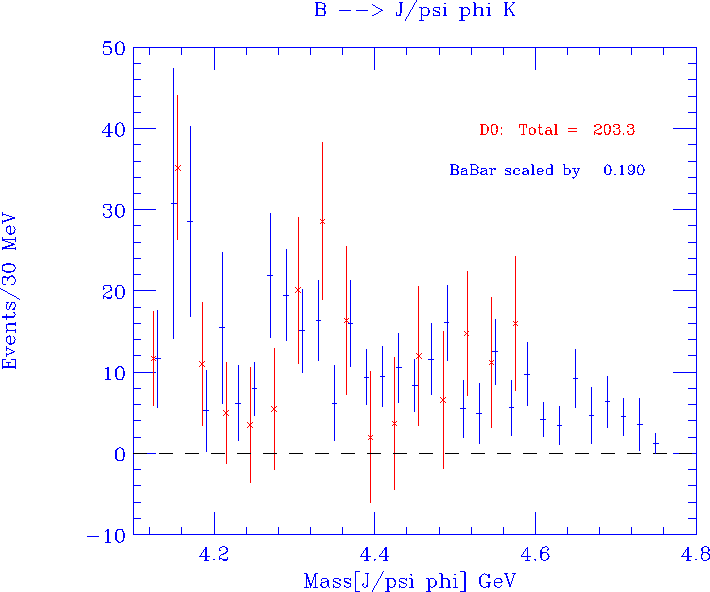}}}\quad
{\scalebox{0.22}{\includegraphics{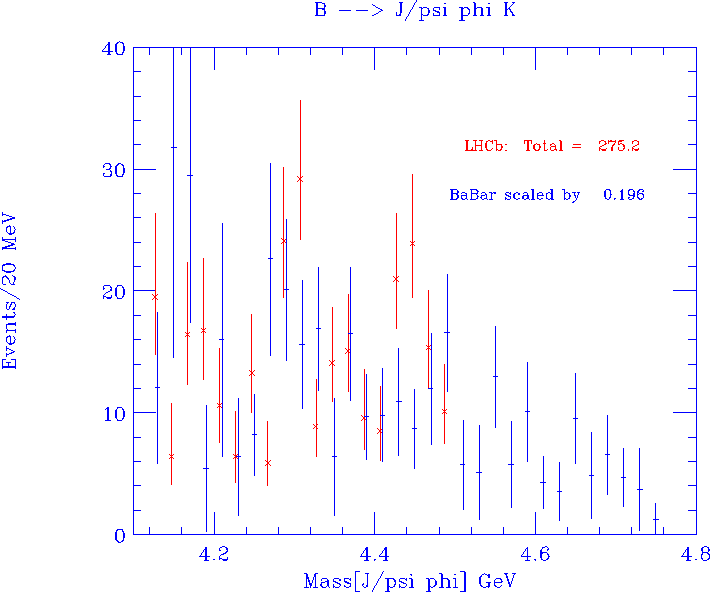}}}\quad
{\scalebox{0.22}{\includegraphics{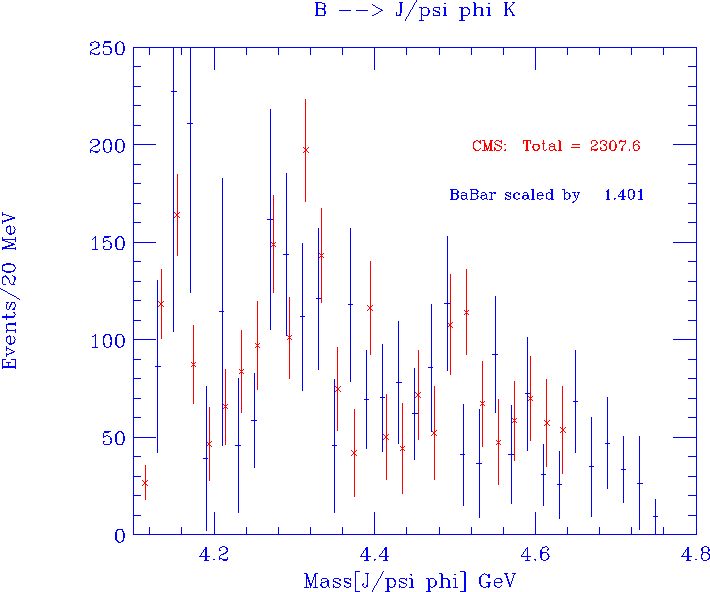}}}
}
\caption{\label{Fig5-prencipe} Invariant mass system of $J/\psi \phi$: comparison between the $BABAR$ data (blue data points), reweighted by the Dalitz efficiency and background-subtracted,  and those published by CDF (top-left), D0 (top-right), LHCb (down-left), and CMS (down-right). Informations are taken from the references ~\cite{kai, LHCb, cms, elisabetta, d0}.}
\end{figure}

As an additional contribution to Ref.~\cite{elisabetta}, which explains in detail this analysis performed by $BABAR$, we provide some plots for a comparison between the $BABAR$ data and the data published in Ref.~\cite{kai,cms, LHCb, d0}. In Fig.~\ref{Fig4-prencipe} you can see a comparison among the data of other experiments that published on the $J/\psi \phi$ invariant mass for $B^+ \rightarrow J/\psi \phi K^+$: the data are scaled by a factor taking in consideration the different integrated luminosity used from each experiment, and they are also background-subtracted, using information from Ref.~\cite{kai,cms, LHCb, d0}. We also rebin properly the data in the histograms of Fig.~\ref{Fig4-prencipe} and ~\ref{Fig5-prencipe}, for the correct comparison. Fig.~\ref{Fig5-prencipe} shows the comparison between the $BABAR$ data, re-weighted by the Dalitz efficiency and background-subtracted,  and what was published from other experiments.
 
The other experiments investigate only the $B^+$ decay within a limited $J/\psi \phi$ mass region, which is different for each experiment, while $BABAR$ analyzed the full range of $J/\psi \phi$ for both decay modes, $B^0$ and $B^+$, where $J/\psi \rightarrow e^+ e^-$ and  $J/\psi \rightarrow \mu^+ \mu^-$. For the other experiments the BFs are not estimated, so we cannot do comparison between our BF measurements and the others.

\begin{figure}[htb]
\centering
\mbox{ 
{\scalebox{0.21}{\includegraphics{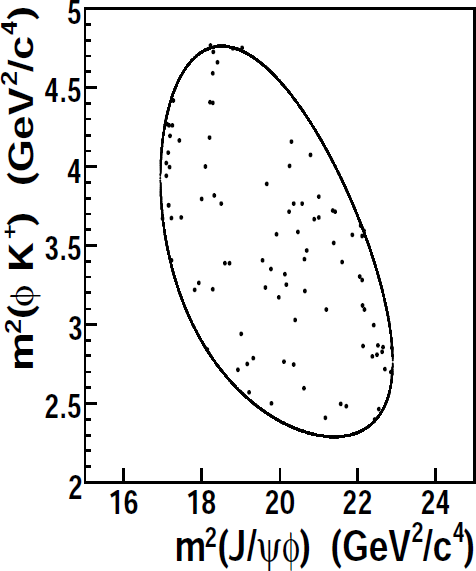}}} \quad
{\scalebox{0.22}{\includegraphics{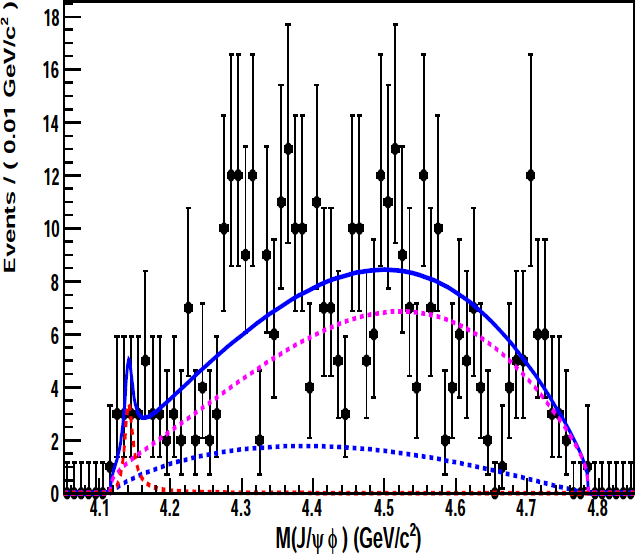}}}\quad 
{\scalebox{0.193}{\includegraphics{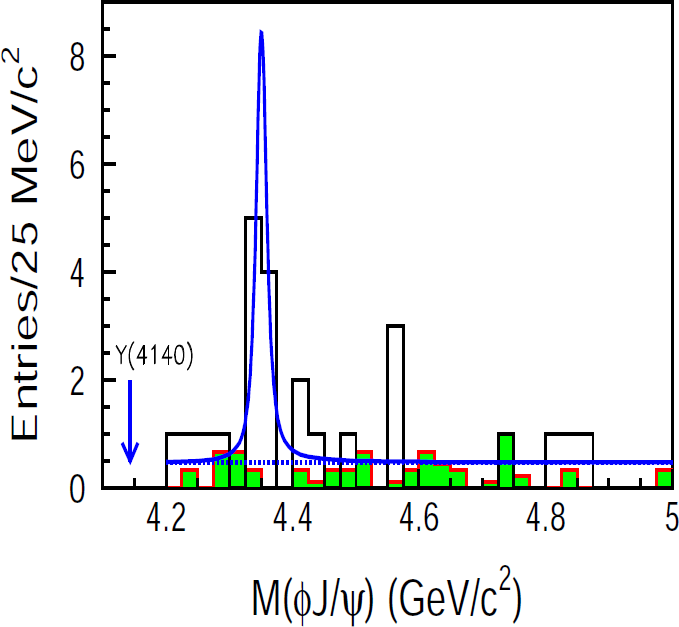}}}
}
\caption{\label{Fig6-prencipe} Dalitz plot of $B^+ \rightarrow J/\psi \phi K^+$ on the Belle data set (left); invariant mass distribution of $J/\psi \phi$ from B decays in Belle (center); invariant mass of $J/\psi \phi$ produced in $\gamma \gamma$ fusion in Belle (right).}
\end{figure}

The Belle experiment has shown the plot of the invariant mass distribution of $J/\psi \phi$ in B decays, but only for $B^+ \rightarrow J/\psi \phi K^+$, as preliminary result\cite{desyLP09}. Belle, having four times larger sample of the reconstructed $B^+ \rightarrow J/\psi \phi K^+$  decay compared to CDF\cite{kai}, has found only $7.5^{+4.9}$$_{-4.4}$ for the Y(4140), from the fit with the resonance parameters fixed to the CDF values (Fig.~\ref{Fig6-prencipe}). As for the $BABAR$ analysis, Belle found low reconstruction efficiency close to the $J/\psi \phi$ mass threshold, as clearly shown in Fig.~\ref{Fig6-prencipe}(center) as dashed magenta curve; so the Belle data have sensitivity lower than the CDF one in that area. While $BABAR$ decided to re-weight the fit function by the 2D-Dalitz efficiency (Fig.~\ref{Fig2-prencipe}), or to re-weight the data directly by the Dalitz efficiency\cite{elisabetta}, event by event, Belle did not perform this study. The upper limit set by Belle, $\cal B$($B^+ \rightarrow Y(4140) K^+) \times$ $\cal B$ $(Y \rightarrow J/\psi \phi) <$6 $\times$10$^{-6}$ remains in agreement with this BF product measured by CDF: $(9.0 \pm 3.4 \pm 2.9) \times 10^{−6}$. The $BABAR$ estimate of the same Belle observable is  $\cal B$($B^+ \rightarrow Y(4140) K^+) \times$ $\cal B$ $(Y \rightarrow J/\psi \phi) <$5.7 $\times$10$^{-6}$, in perfect agreement with the Belle result.
Belle has also investigated the $J/\psi \phi$ mass system produced in the $\gamma \gamma$ fusion processes\cite{ref3}. In this reaction, states with J$^{PC}$=0$^{++}$ or 2$^{++}$ are allowed to be formed. In the analysis performed by Belle, no evidence of signal is found at a mass value m = 4.14 GeV/c$^2$, as shown in Fig.~\ref{Fig6-prencipe}(c); indeed, a narrow state is seen around m = 4.35 GeV/c$^2$. Belle measured upper limits on the product of the two-photon decay width of the $Y(4140)$ and its BF: $\Gamma_{\gamma \gamma} (Y(4140))$$\cal B$$(Y(4140) \rightarrow J/\psi \phi) <$ 41 eV for $J^P$ = 0$^+$ or $<$ 6.0 eV for $J^P$ = 2$^+$. These values are much lower than predicted for Y(4140), in the hypothesis that it would be a $D_S^{*+}D_S^{*-}$ molecule.

In summary, both $BABAR$ and Belle have not find evidence for the two resonances claimed by the CDF collaboration, but they could evaluate upper limits on the presence of these states in the invariant mass system of $J/\psi \phi$. Both $BABAR$ an Belle experiments show lower sensistivity at threshold of the invariant mass distribution of $J/\psi \phi$ due to the poorer reconstruction of kaons. $BABAR$ has corrected data by the efficiency, evaluated from 2D-Dalitz plot study on MC PHSP simulations.  Systematic effect due to polarization are taken into account, as explained in Ref.~\cite{elisabetta}. 

\end{document}